# Electronic correlations, Jahn-Teller distortions and Mott transition to superconductivity in alkali-$C_{60}$ compounds

H. Alloul

Laboratoire de Physique des Solides, UMR 8502 CNRS, Université Paris-Sud 91405, Orsay (France)

**Abstract.** The discovery in 1991 of high temperature superconductivity (SC) in $A_3C_{60}$ compounds, where $A$ is an alkali ion, has been rapidly ascribed to a BCS mechanism, in which the pairing is mediated by on ball optical phonon modes. While this has lead to consider that electronic correlations were not important in these compounds, further studies of various $A_nC_{60}$ with n=1, 2, 4 allowed to evidence that their electronic properties cannot be explained by a simple progressive band filling of the $C_{60}$ six-fold degenerate $t_{1u}$ molecular level. This could only be ascribed to the simultaneous influence of electron correlations and Jahn-Teller Distortions (JTD) of the $C_{60}$ ball, which energetically favour evenly charged $C_{60}$ molecules. This is underlined by the recent discovery of two expanded fulleride $Cs_3C_{60}$ isomeric phases which are Mott insulators at ambient pressure. Both phases undergo a pressure induced first order Mott transition to SC with a ($p, T$) phase diagram displaying a dome shaped SC, a common situation encountered nowadays in correlated electron systems. NMR experiments allowed us to study the magnetic properties of the Mott phases and to evidence clear deviations from BCS expectations near the Mott transition. So, although SC involves an electron-phonon mechanism, the incidence of electron correlations has an importance on the electronic properties, as had been anticipated from DMFT calculations.

## 1 Introduction

The discovery in 1985 of new pure carbon molecular states, in the form of quasi spherical or ellipsoidal structures, called Buckminster fullerenes [1], have been at the origin of new materials in condensed matter physics. The most remarkable of these molecules is certainly $C_{60}$, which has the shape of a soccer ball. It could soon be produced in macroscopic quantities [2] and isolated, yielding then the production of pure solid $C_{60}$. In this compound (figure 1 a), the voids between the $C_{60}$ balls are large enough to house a variety of ions. Thus, the insertion of alkali atoms which donate their $s$ electron to the $C_{60}$ balls yields a series of $A_nC_{60}$ compounds which have been extensively studied as they display remarkable physical properties, for instance high temperature superconductivity in $A_3C_{60}$ compounds [3].

The $C_{60}$ compounds are plastic crystals at high temperature, as the $C_{60}$ balls undergo fast molecular reorientations and can be considered as simple spheres. The solid usually has then fcc or bct structure depending of the size and number of inserted ions. For instance (figure 1), both pure $C_{60}$ and $A_3C_{60}$ solids are fcc (space group *Fm3m*). But when $T$ is lowered the $C_{60}$ molecular rotations freeze out progressively [4-5] yielding eventually a structural change, as for pure $C_{60}$ which becomes simple cubic with *Pa*3 space group below 263K.

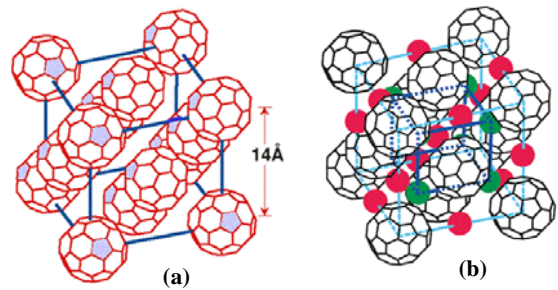

**Fig. 1.** Fcc structures (a) of $C_{60}$ and (b) of $A_3C_{60}$. There, the two tetrahedral and the octahedral alkali sites are differentiated. The disorientations of the pentagon units in (a) illustrate the low temperature orientational disorder of the $C_{60}$.

In some cases, the dynamics of the orientations of the fullerene molecules plays a crucial role in the physical properties of alkali fulleride materials. The most spectacular example is certainly the formation of one-dimensional polymeric chains by 2+2 cyclo-additional solid state reaction in the $A_1C_{60}$ crystals [6]. The rate of reaction is governed by the fraction of neighbouring $C_{60}$ molecules with relative orientations appropriate for the cyclo-addition to occur. This polymer structure could be evidenced by NMR [7] and x-rays [8] and has been found to display one dimensional metallic [9] and magnetic [10] properties.

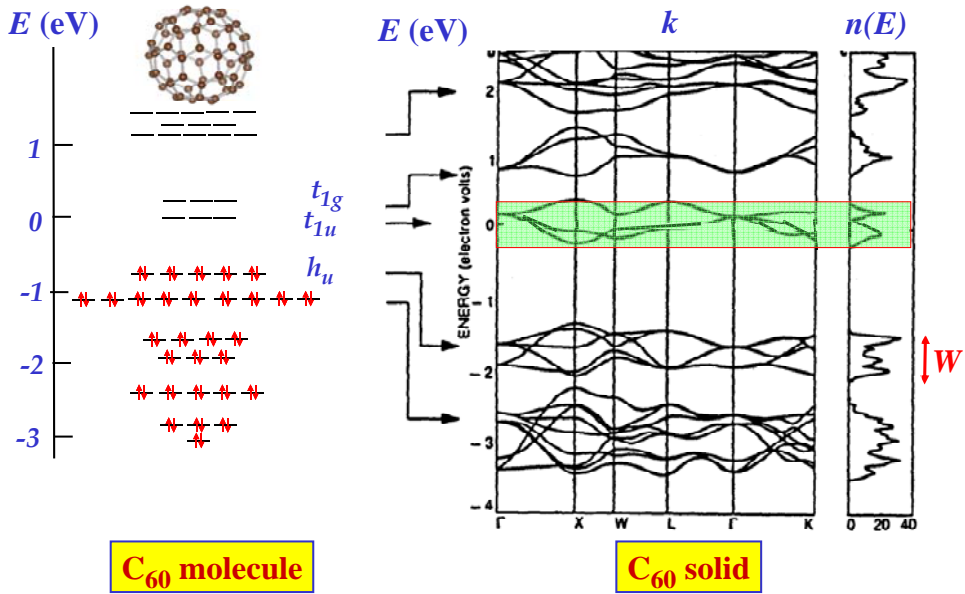

**Fig. 2.** The π electronic molecular energy states of C$_{60}$ shown on the left give narrow bands of width $W$ ~0.5eV in the solid. In pure C$_{60}$ the $h_u$ band is filled. Upon alkali insertion the calculated LDA bands are not markedly modified. The $s$ levels of the alkali being much higher in energy, the alkali valence electrons are fully transferred to the C$_{60}$ molecular bands, and progressively fill the lowest unoccupied $t_{1u}$ band for increasing $n$ in $A_nC_{60}$ (adapted from ref.[22-23]).

In most $A_nC_{60}$ the freezing of the rotations at low $T$ retains a substantial amount of orientational disorder, as for instance in fcc $A_3C_{60}$ [11]. In those compounds one alkali atom per unit cell is inserted in an octahedral site, while two others are at the tetrahedral site of the fcc structure at high $T$ (figure 1 b). A direct manifestation of the low $T$ merohedral disorder is the observation by NMR of two distinct signals for the tetrahedral alkali site NMR [12], which are associated to distinct local orders of the C$_{60}$ surrounding the tetrahedral site [13].

Insertion of alkali atoms is a well known process to induce metallic behaviour in carbon based solids, such as graphite [14]. In fullerides, the charge transfer from the alkali atoms to the fullerene molecules is nearly complete, and the electrons transferred from the alkalis to the C$_{60}$ balls are expected to participate in delocalized metallic states. We shall recall here that the weak transfer integrals between C$_{60}$ molecules imply electronic structures with small bandwidths, so that electronic correlations are important [15].

Furthermore adding electrons on the C$_{60}$ balls cannot be done without taking into consideration the coupling of the molecular electronic states with the inner vibration modes of the molecule, which induce Jahn-Teller effects [16]. This yields the observation of a series of effects associated with molecular physics and strong correlations in these compounds.

However the properties of $A_3C_{60}$ compounds have been initially considered as weakly correlated metals displaying superconducting BCS behaviour [17], until the recent observation that expanded states are indeed strongly correlated [18], and even exhibit a Mott transition from an insulating magnetic state to a metallic and SC state when the distance between C$_{60}$ is decreased by an applied pressure [19] [20]. Those results confirm anticipations from calculations of the ground state energies assuming strong correlations and Jahn-Teller effects [21].

So we recall in section 2 the electronic structure of the molecule and that expected for the solid in the band approximation. We discuss then in section 3 the expected incidence of electron phonon coupling and Jahn-Teller effects for charged molecules. In section 4 we present the case of CsC$_{60}$, for which the observation of charge segregation demonstrates the stability of doubly charged molecules. This will lead us to explain in section 5 why evenly charged solids are not metallic and display Mott Jahn-Teller insulating states. Finally in section 6 we shall consider the $A_3C_{60}$ compounds and their superconducting properties and give evidence that in expanded cases a Mott transition to SC reveals that electronic correlations are indeed important in these compounds.

## 2 From the C$_{60}$ molecule to doped solids

The bonding between C atoms in C$_{60}$ molecules is, as for the graphene sheet, mediated by the $sp^2$ σ bonds between neighbouring C atoms, the difference being that C pentagons are required here to establish the curvature of the surface. These bonds house three of the four $2s^2p^2$ electrons of the C atoms. The 60 remaining $2p$ electrons participate in delocalized π electrons similar to those of the benzene ring. These π molecular orbitals are filled by 60 electrons and the corresponding highest occupied $h_u$ molecular orbitals (HOMO) are separated from the lower unoccupied $t_{1u}$ orbitals (LUMO) [22], which are triply degenerate and can therefore house up to 6 electrons per C$_{60}$ (figure 2). Those $t_{1u}$ molecular orbitals play, for the $C_{60}$ molecule, a role quite analogous to that of the three $p$ orbitals $p_x$, $p_y$, $p_z$ for a carbon atom.

In the solid, the $C_{60}$ molecules being at distances comparable to the van der Waals distances, the transfer integrals between the molecular states are rather weak. The resulting bands have narrow widths of about 0.5eV, so, as can be seen in figure 2, the bands arising from distinct molecular levels are not interleaved [23].

For $A_nC_{60}$ compounds, the alkali $s$ levels are higher in energy than the LUMO states within the local density approximation (LDA). Therefore the band structure is quite similar to that of $C_{60}$. The electrons donated by the alkali can then be considered as dopants which shift the Fermi level through the $t_{1u}$ band up to $n=6$. Within this rigid band filling approach all compounds with $n$ distinct from 0 or 6 are then expected to be metallic.

## 3 Jahn–Teller effect in charged $C_{60}^{n-}$

In such calculations one assumes that adding electrons on the $C_{60}$ balls does not change the molecular energy levels. This is clearly not valid, as modifying the charge of the $C_{60}$ ball is expected to lift the degeneracy of the levels by the well known Jahn-Teller effect [24]. Usually in a solid this occurs by a spontaneous lattice distortion, which reduces the symmetry at the origin of the degeneracy. Here, for a charged molecule, a static or dynamic distortion of the molecule occurs and, as the electrons only occupy the low lying states, this corresponds to an energy gain with respect to the non distorted case.

The magnitude of the energy gain, and the symmetry of the distortions have been studied in some detail [16-17] by considering the coupling of the molecular orbitals with the phonon modes of the $C_{60}$ balls. Two possibilities are anticipated, depending of the symmetry kept for the $C_{60}$ ball. An axial elongation of the molecule splits the three levels in a lower singlet and a doublet for singly or doubly charged molecule. For one or two holes in the $t_{1u}$ level, the ball is squeezed and the levels are inverted. For three electrons on the ball all symmetries are lost which gives rise to three independent levels.

We present then, for pedagogical purpose in figure 3 a simplified sketch of this ordering of the $t_{1u}$ levels as a function of the $C_{60}$ charge. There the calculated level splitting of the order of 0.1eV do not take into account the change of the coulomb interactions between electrons which has comparable magnitude.

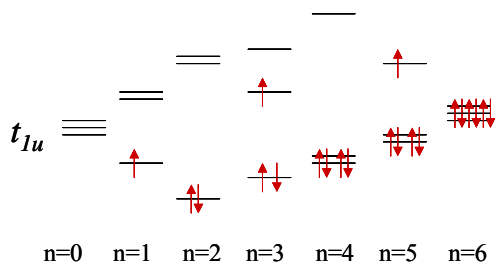

**Fig. 3.** Representation of the modifications of the electronic energy levels induced by the JTD as a function of $-ne$, the **$C_{60}$** charge. Here interactions between electrons are not considered.

One would like of course to detect these JTD by structural observations. This would only be possible if the distortion is static, while the configuration of the ball could fluctuate between different spatial orientations of the distortion. In some molecular salts, in which charged states are spatially well separated, static JTD could be evidenced [25]. The C atom radial displacements were found at most of the order of 0.05A. The possible dynamic nature of the JTD and the small magnitude of the spatial distortion explain the difficulty to detect them in the alkali fullerides. In that case, we shall however see in the next sections that the corresponding modifications of molecular energy levels have profound consequences on the electronic properties.

## 4 $A_1C_{60}$ compounds: local singlets

It is certainly for $n=1$ that one expects to detect most easily the incidence of electronic correlations in fullerides. In these phases the alkali occupies at high $T$ the octahedral site of the fcc arrangement of the $C_{60}$ balls (figure 4 top). As we have indicated in the introduction, these compounds do however display a complicated phase diagram, as a structural transition toward a polymer phase occurs below 300K, so that the fcc phase could initially only be studied at the highest temperatures.

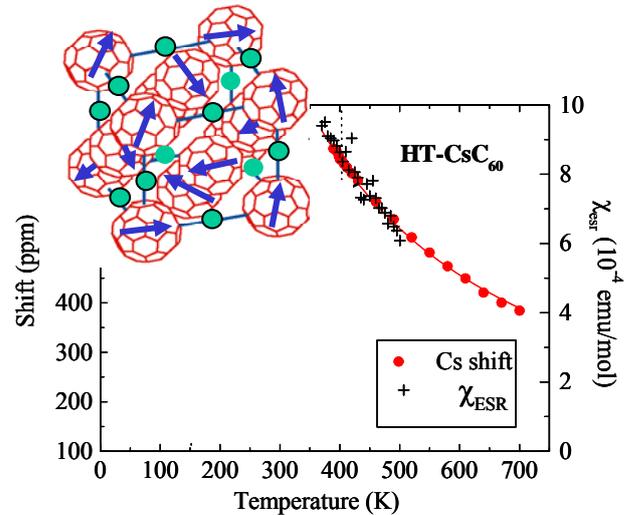

**Fig. 4.** In the high $T$ $CsC_{60}$ phase, the Cs fill the octahedral sites of the $C_{60}$ fcc structure. The $^{133}$Cs NMR shift [26] and the ESR signal intensity [27] both display a Curie type variation, as shown by the full line fit [28]. Hence the disordered arrows on the $C_{60}$ balls, which sketch the $S=1/2$ Mott insulating state.

There, both NMR [26] and ESR studies [27] indicate that the spin susceptibility displays a Curie law, with an effective moment which corresponds to a spin $S=1/2$ and a very weak extrapolated Weiss temperature [28], as can be seen in figure 4. Such a Mott insulating state is somewhat expected and indicates that the electrons donated by Cs are localized on the $C_{60}$. One could have been then able to study the insulator to metal transition under an applied pressure if the system did not transform into a polymer phase.

One could however succeed in avoiding the polymerization by quenching the compound at low $T$, which retains the cubic $Pa3$ phase with orientational order, like pure $C_{60}$ [29]. This phase is found to transform above 130K in a semiconducting phase in which the $C_{60}$ are dimerized. This permitted us then to study by NMR [30, 31] the cubic quenched phase below 130K, and to reveal its original properties that we shall recall below, which provide a nice illustration of the role of Jahn-Teller related effects in fullerides.

Our NMR data revealed that, while a single octahedral Cs crystallographic site occurs in this cubic quenched phase, the NMR spectra display three $^{133}$Cs lines, two of them of comparable intensities while the third one only corresponds to about 6% of the Cs sites (figure 5). These lines are well separated at the highest temperatures and we labelled them in accordance with their position with respect to an ionic Cs reference as NS (Non Shifted), S (Shifted) and 2S (twice Shifted).

Obviously this observation indicated that the electronic environment of these Cs sites differ markedly. Special double resonance NMR techniques [30] allowed us to demonstrate that these Cs sites are intermixed at the microscopic scale, and do pertain to a single phase. We have shown in ref. [31] that this spectrum can be very accurately understood by the model presented in figure 5, in which some $C_{60}$ balls are doubly charged, the two electrons being stabilized in a singlet state at $T=0$. The NS, S and 2S lines arise then from Cs nuclei with respectively 0, 1 or 2 near neighbor $C_{60}^{2-}$ balls. This charge segregation is static on the NMR timescale (msec). From the relative intensities of the lines, we deduced that singlets occur on about 12% of the $C_{60}$.

The NMR shifts of the $^{133}$Cs sites allowed us to probe the electronic susceptibility of their neighbouring $C_{60}$ balls. While the NMR shift of the NS site is found $T$ independent, as in a metallic state, that of the S and 2S sites exhibit a large increase at high $T$ as displayed in figure 6. Such an increase of the electronic susceptibility of the singlet is attributed to the thermal population of its triplet excited state. A similar differentiation is found for the spin-lattice relaxation rates $1/T_1$ of the three lines, with a small constant $(T_1T)^{-1}$ for the NS line and a large increase versus $T$ for the S and NS lines.

Let us point out that, in view of the relatively small number of singlets, it is not surprising to notice that the macroscopic probes, such as the susceptibility or equivalently the ESR intensity [32], are found nearly $T$ independent. This metallic like behavior is that probed by the NS site, and corresponds to that of the dominant fraction of $C_{60}$ balls which bear less than one electron per ball. So this localization of electronic singlets produces altogether a decrease of the electronic charge of the remaining $C_{60}$. It is then natural to find that such a hole doping has suppressed the Mott Hubbard insulating behaviour of the high temperature fcc phase, and that the metallic spin susceptibility does not extrapolate to the Curie Weiss law of the high $T$ cubic phase.

So although the strength of the Coulomb repulsion a priori forbids the double occupancy of a $C_{60}$ ball, the Jahn-Teller interaction, which has the opposite tendency, favors evenly charged $C_{60}$, as explained in sec. 3. This charge segregation is certainly helped as well by the decrease of band electronic energy due to the electron delocalization of the remaining electrons.

This observation in the cubic quenched phase of $CsC_{60}$ is therefore an important result, as it proves that *Jahn-Teller interactions are important enough to oppose Coulomb repulsion in some cases*. Furthermore this experiment also demonstrates that the doubly charged molecule is in a low spin $S=0$ singlet state and that the triplet state is higher in energy, so that the Jahn-Teller distorted state violates the ordering usually expected from Hund's rules.

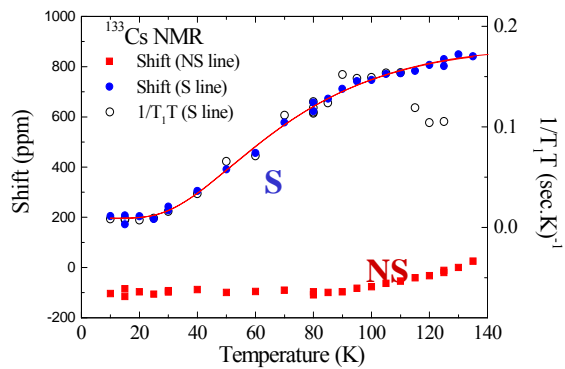

**Fig. 6.** $T$ dependences of the NMR shifts of the $^{133}$Cs NS and S lines. The right scale for the $(T_1T)^{-1}$ data has been chosen to fit that of the NMR shift for the S line [31].

## 5 Evenly charged compounds: Mott Jahn-Teller insulators

Another case which has led to consider in some detail the role of JTD has been that of evenly charged compounds which are expected to be metallic from band calculations but are found to be insulating. Their optical conductivity indeed reveals the absence of the metallic Drude peak [33]. Electron energy loss spectroscopy experiments [34], as those displayed in figure 7 allow to evidence gaps ~0.5eV in the optical spectra. There, the peak detected at about 1eV in all compounds corresponds to the first interband transition from $t_{1u}$ to $t_{1g}$ expected from the band structure (see figure 2).

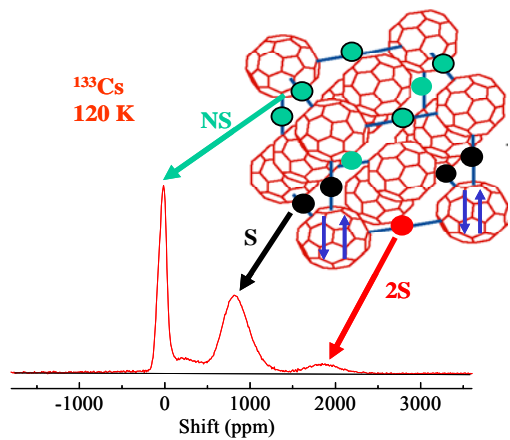

**Fig. 5.** The $^{133}$Cs NMR spectra [30] in the cubic quenched phase of $CsC_{60}$ exhibit three lines NS, S and 2S (see text). This reveals a differentiation of the electronic charges of their neighbouring $C_{60}$, represented on the top part of the figure, with the localisation of about 12% of doubly charged $C_{60}$.

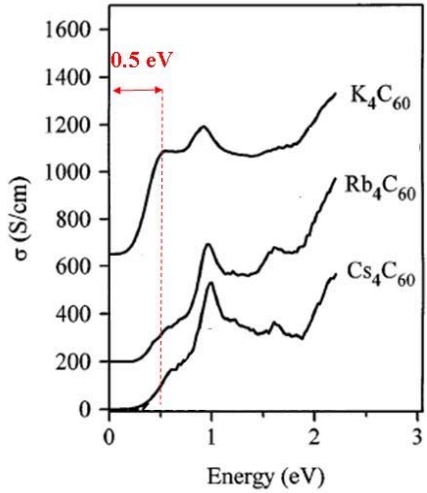

**Fig. 7.** Optical conductivity from electron energy loss spectroscopy measurements performed in various $A_4C_{60}$ compounds. They exhibit ~0.5eV gaps (adapted from [34]).

The existence of small optical gaps might naively be expected if the degeneracy of the $t_{1u}$ level is lifted by a specific ordering of the $C_{60}$ balls in the structure. However the insertion of four large alkali ions such as K, Rb or Cs in $A_4C_{60}$ forces a distortion of the fcc structure of pure $C_{60}$ into a bct structure. On the contrary $A_2C_{60}$ compounds are quite difficult to stabilize as the alkali ions are preferentially filling the large octahedral void which they occupy in the $A_1C_{60}$ compounds. The only known compound with two electrons per ball is $Na_2C_{60}$, for which the small Na alkali ions are stable in the tetrahedral sites. In this compound even the low temperature cubic orientational order of pure $C_{60}$ is preserved. So the fact that $Na_2C_{60}$ and $K_4C_{60}$ display similar optical gaps is a hint against an explanation based on structural effects.

Of course, as pointed out in section 3, the degeneracy of the molecular level can as well be lifted by a Jahn-Teller distortion of the charged molecule, which would explain the 0.5eV gap magnitude. However with the similar magnitude of the transfer integrals between $C_{60}$, the substructure should be washed out by the band effects and a metallic state would then result. So JTD alone cannot explain the insulating state.

Another possibility is of course then to consider that electron correlations are strong enough to localize the two or four electrons on the $C_{60}$, so that these systems would be Mott insulators. But here-again, if one considers strong correlations alone, the electronic states on a ball should have a high spin configuration according to Hund's rules. Here NMR observations are quite conclusive, as the $^{13}C$ NMR signal remains quite narrow and is not shifted at low temperature. This absence of static local fields on the $^{13}C$ establishes then that the insulating ground state is non magnetic [35]. So, as has been anticipated by Fabrizio et al [36], a low spin Jahn-Teller distorted molecular configuration appears to be stabilized by the electronic interactions. Those contribute to localize the even number of electrons on the balls. Such an electronic structure has been named as a *Mott Jahn Teller insulating state*.

NMR measurements of the spin lattice relaxation of $^{13}C$ do allow to probe the excited states of this electronic structure [35], and give evidence for the existence of gaps of 50 meV in $Na_2C_{60}$ and 140mev in $K_4C_{60}$ (figure 8), in agreement with other magnetic measurements [37,38]. Such gaps, much smaller than the measured optical gaps, confirm then that we are not dealing with a simple semiconductor like band structure effect. The nuclei are probing directly here a small gap in *the spin excitations of the JTD state*. This means that the spin gap detected by NMR is indeed the difference in energy between the singlet and the excited triplet state which corresponds to a distinct JTD, a situation quite reminiscent of that found in §4 for the localized singlets in the $AC_{60}$ compounds.

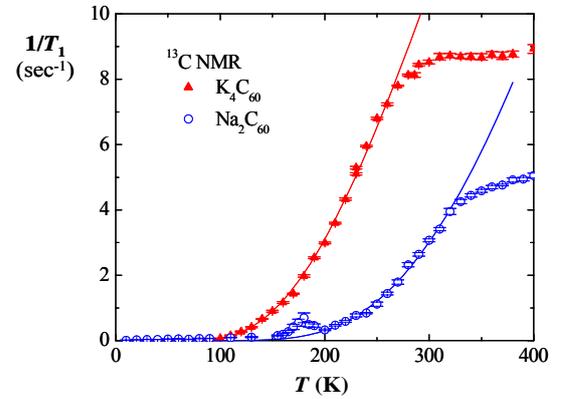

**Fig. 8.** Variation of the $^{13}C$ NMR relaxation rates $1/T_1$ in $Na_2C_{60}$ and $K_4C_{60}$ versus $T$ [35]. The activated behaviour corresponds to gap energies respectively of 60 meV and 140 meV (full lines). The small peak in $1/T_1$ detected at 180K in $Na_2C_{60}$ is an excess nuclear spin relaxation rate due to the motion of the $C_{60}$ at the orientational ordering transition. The strong deviation from the activated law detected at high $T$ in both compounds might correspond to a change of electronic structure.

Finally this Mott JT insulating state is not so far from a metallic state, the borderline cases being those with the smallest inter-ball distance. A very small residual $T$ linear nuclear spin lattice relaxation rate is indeed detected at low $T$ in $Na_2C_{60}$ [39]. Such a linear dependence corresponds to the existence of a small density of states at the Fermi level. In the case of $Rb_4C_{60}$, its appearance under an applied pressure [40] indicated that an insulator to metal transition can be triggered by a reduction of the inter-ball distance, a situation that we shall encounter later as well in the $n=3$ compounds.

To conclude this section, we have seen here that the detected optical gap is attributed to the direct gap opened by the JTD between the $t_{1u}$ levels, while the spin-gap corresponds to a singlet-triplet transition between two different JTD states as discussed in section 3. This interpretation is supported by the very fact that the magnitudes of these two gaps correspond very well to those expected in the JTD framework.

A direct evidence for a static JTD in this phase has however not been established as no superstructure could be detected so far, even by extensive x-ray investigations, done for instance in $Rb_4C_{60}$ [41], so that the JTD is more likely dynamic. Alternatively, infrared measurements allowed to detect the symmetry of the vibrational modes in $K_4C_{60}$ [33][42], and evidenced a change at 250K which could be connected with the electronic structure change seen in $1/T_1$ in figure 8, which remains to be understood.

# 6 Superconductivity and magnetism in $A_3C_{60}$ compounds

We have seen here-above that the interactions between electrons and JTD have major influence on the electronic properties of some alkali fullerene compounds. One could think then that this is the case as well for compounds with three electrons per ball or $A_3C_{60}$, which are usually metallic and superconducting [17], with reported critical temperatures as high as ~40K [43]. We shall describe here first the experimental results which have led a majority of researchers to consider initially that this was not the case, and that superconductivity appeared purely driven by an electron-phonon interaction in the BCS weak coupling limit (§6.1). We will then show in §6.2 that this shortcut was not fully justified, and that electron correlations ought to be considered. Finally, we shall introduce in §6.3 the new experiments on the recently discovered expanded isomeric $Cs_3C_{60}$ compounds which undergo a Mott insulator to SC transition under applied pressure. Those definitely confirm the importance of correlations, and open a new field of investigations.

### 6.1 BCS superconductivity in $A_3C_{60}$?

Contrary to the evenly charged fullerides, the $A_3C_{60}$ compounds are usually metallic and present a regular Drude peak in their optical conductivity [44]. Also the normal state electronic susceptibility, as measured by the electron spin resonance (ESR) intensity is practically $T$ independent [45], that is Pauli like, as confirmed as well by NMR Knight shift and $T_1T$ data [46]. Quantitatively the effective mass enhancement is $m^*/m_0$ ~3, so that one is tempted to conclude that these compounds are not far from regular Fermi liquids [17,47].

In the superconducting state, a jump of the specific heat is detected and increases with $T_c$ as expected for a BCS transition [48]. Similarly, the SC gap $2\Delta$ scales with $T_c$, with a $\Delta/k_BT_c$ ratio not far from the 3.5 value expected in the theoretical BCS scenario [17]. NMR in the superconducting state shows up the decrease of Knight shift, that is the loss of spin susceptibility, expected for singlet pairing [49,50]. Finally Hebel-Slichter coherence peaks in both the NMR $(T_1)^{-1}$ [49,50] and the muon spin relaxation rate [51] have been observed below $T_c$ in weak applied fields, which also points towards *conventional s-wave symmetry of the order parameter.*

But, among the experimental observations, the most emphasized one has been the variation of $T_c$ with the chemical nature of the inserted alkali, which permits to change the inter-ball distance [17]. One could monitor, as shown in figure 9, a continuous decrease of $T_c$ with lattice compression in fcc $A_3C_{60}$. There, the detected $T_c$ variation is the same for chemical [52] and pressure changes [53] of the lattice parameter. This absence of alkali mass isotope effect has been initially assigned to a BCS weak coupling relation for $T_c$

$$k_B T_c = \hbar\omega_D \exp[-1/VN(E_F)], \qquad (1)$$

in which the density of states $N(E_F)$ appears to be the single driving parameter. This has led to consider an on ball electron-phonon coupling $V$, the active phonons in $\hbar\omega_D$ being then the high energy optical phonons ($H_g$), which are responsible for the JT effects [17] [54-56]. However, it has been pointed out that, as displayed in figure 9, this relation does not extend to samples of $Na_2AC_{60}$, which order in the cubic $Pa3$ structure [57,58], though a similar law might hold with a larger slope [59].

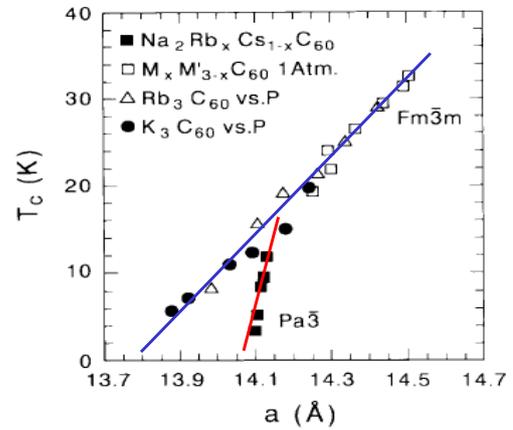

**Fig. 9.** $T_c$ variation versus fcc lattice parameter obtained by chemical exchange of the alkali (M /M') or by applying a pressure which leads to a decrease of DOS and of $T_c$. The samples with cubic $Pa3$ structure do not follow the same trend as those with $Fm3m$ structure (adapted from [59]).

This difference has raised questions about the possible incidence of the structure, of the nature of the alkali and of the merohedral disorder on the density of states [60] which has to be considered in equ.(1). Actually, it has been shown by ARPES that the arrangement of the molecules on deposited single layer fulleride films at the half filled composition does have a dominant influence on the electronic properties of the films [61].

### 6.2 Incidence of JTD and electronic correlations

Considering these results one is led to question whether the correlation effects and JTD which induce the Mott insulating state of the evenly charged compounds do play any role in the electronic properties of these $A_3C_{60}$ compounds. We have seen in section 5 that the Mott JT states for even $n$ are already not so far from a metallic state. Does that mean that odd parity of $n$ would indirectly favor hopping rather than localization?

To answer that question, one might consider the energy cost of transferring one electron on a nearby site,

that is transforming $2C_{60}^{-n}$ into $C_{60}^{-(n+1)}$ and $C_{60}^{-(n-1)}$. If $E(n)$ is the molecular energy of a ball with $n$ electrons, this energy cost is given by

$$U_{eff} = E(n+1) + E(n-1) - 2E(n). \quad (2)$$

So an examination of figure 3 leads one to conclude that the contribution of the JTD to $U_{eff}$ will be positive for even $n$, and negative for odd $n$. The JTD stabilisation of evenly charged $C_{60}$ therefore adds to the on ball coulomb repulsion for even $n$ and reduces it for odd $n$ [62-63]. This explains the occurrence of large correlation effects and of the JTD Mott insulating state in the evenly charged compounds. But conversely this can be taken as an argument for smaller overall electronic correlations in the $A_3C_{60}$ compounds. However the magnitude of $U_{eff}<0.5$ eV, as compared to $U \sim 1$ eV, would hardly justify a total loss of electronic correlations.

Various efforts have then been undertaken to estimate the incidence of electronic correlations on electron phonon superconductivity in the case of these $A_3C_{60}$. Usually, the Coulomb electronic interactions are taken into account within the McMillan extension of BCS through an empirical parameter $\mu^*$

$$k_B T_c = \hbar \omega_D \exp [ -1/(\lambda - \mu^*)], \quad (3)$$

where $\lambda = VN(E_F)$. This equation applies when $\hbar \omega_D \ll E_F$ while here the phonons, the JTD and the Fermi energies have the same order of magnitude, which does not allow a standard use of this approach[a]. More recent calculations done within Dynamical Mean Field Theory (DMFT), suggest that in this specific case of Jahn-Teller on ball phonons, the electronic correlations would not be very detrimental to phonon driven superconductivity [64].

In any case a different school of thought has been suggesting for long that electron-electron correlations might even drive superconductivity in such compounds [15-16]. It has even been proposed that $A_nC_{60}$ should be Mott insulators and are only metallic due to a non stoichiometry of the alkalis [65]. Although our result for the charge segregation in the cubic quenched phase of $CsC_{60}$ could be supportive of this possibility, no experimental evidence along this line could be found so far in $A_nC_{60}$. It has rather been seen that $T_c$ peaks at $n=3$ if one varies the alkali doping [66].

To better qualify the proximity to a Mott insulating state, attempts to produce expanded fulleride compounds have been done for long. A successful route has been to insert $NH_3$ neutral molecules to expand the $A_3C_{60}$ lattice. It was found that $(NH_3)K_3C_{60}$ is insulating at ambient pressure [67], and becomes SC with $T_c$ up to 28K under pressure [68]. Furthermore at 1bar it was shown to be an AF [18]. However this insertion of $NH_3$ induces an orthorhombic distortion of the lattice which is retained, though modified, under applied pressure [69]. So, although this system undergoes a Mott transition to a metallic state, this occurs in an electronic structure where the $t_{1u}$ level degeneracy has been lifted by the specific spatial order of the $C_{60}$ balls. Let us recall then that, for a single orbital case the Mott transition occurs for $U/W=1$, while for $N$ degenerate orbitals the critical value $U_c$ is expected to be larger [62], typically $U_c=N^{1/2}W$. Therefore $U_c$ is reduced in the case of $(NH_3)K_3C_{60}$ with respect to that of $A_3C_{60}$, for which the $t_{1u}$ orbital degeneracy is preserved. Though this complicates any direct comparison between these two cases, these results establish that $A_3C_{60}$ are not far from a Mott transition, and hence that electronic correlations cannot be neglected.

### 6.3 Mott transition in $Cs_3C_{60}$ isomeric compounds

A much simpler way to expand the $C_{60}$ fcc lattice has been of course to attempt the insertion of Cs, the largest alkali ion, by the synthesis procedure used for the other $A_3C_{60}$. But, the bct $Cs_4C_{60}$ and the polymer phase $Cs_1C_{60}$ being stable at room $T$ (see § 1), the $Cs_3C_{60}$ composition is found metastable. In a mixed composition sample it was found that a $Cs_3C_{60}$ phase becomes SC at high pressure with $T_c \sim 40$ K [43]. A distinct chemical route finally recently permitted to produce samples containing the fcc-$Cs_3C_{60}$, mixed with an A15 isomeric phase. The latter has been found insulating at ambient pressure, and becomes SC under pressure, exhibiting a SC dome with a maximum $T_c$ of 38K at about 10kbar [19].

In similar mixed phase samples, NMR experiments allowed us to separate the specific $^{133}Cs$ spectra of the two isomers [20]. As displayed in Fig 10, the A15 phase has a single Cs site with non cubic local symmetry, hence its seven lines quadrupole split spectrum ($^{133}Cs$ has a nuclear spin I=7/2). The fcc phase spectra exhibit the usual features found for all other fcc-$A_3C_{60}$, displaying the signals of the orthorhombic (O) site and of two tetrahedral sites (T, T'). Those are differentiated by the local orientation orderings of the $C_{60}$ around the alkali, associated with the merohedral disorder (see § 1).

This possibility to select the signals of the two phases allowed us to demonstrate that both are in a paramagnetic insulating state at $p$=1bar, and display a Mott transition at distinct applied pressures [20].

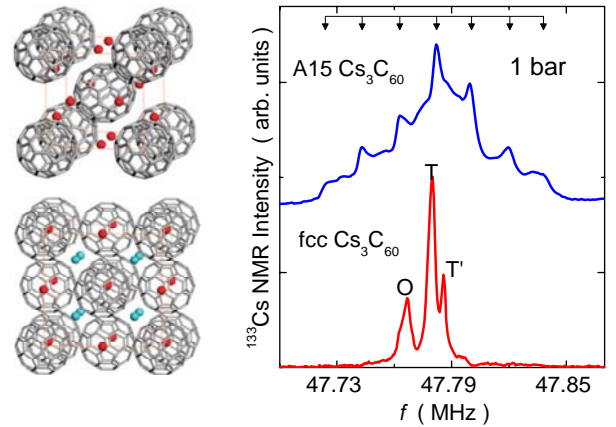

**Fig. 10.** The unit cells of the A15 and fcc isomeric phases of $Cs_3C_{60}$ are displayed on the left [19]. Their $^{133}Cs$ NMR spectra [20] taken at 1bar and $T$=300 K are reported on the right. The specific features of the spectra are discussed in the text.

---

[a] Using the more accurate Migdal-Eliashberg equation does not change this conclusion, see [17].

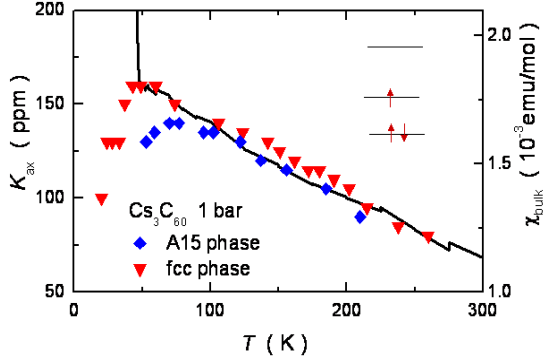

**Fig. 11.** Paramagnetic behaviour detected by SQUID data (right scale) and from the anisotropic shift $K_{ax}$ of the $^{13}$C NMR in the two isomeric Cs$_3$C$_{60}$ phases [20]. No significant difference can be detected above 100K and the effective moment corresponds to the low spin JTD state sketched in insert. The magnetization increase below 47K is that of the A15 phase Néel state.

We could further study their magnetic behaviour at ambient pressure using data on the $^{13}$C nuclear spins, which are directly coupled to the electron spins of the $t_{1u}$ molecular orbitals. The coupling being dipolar [50], it induces a characteristic anisotropic broadening of the $^{13}$C NMR which allowed us to monitor the C$_{60}^{3-}$ magnetism.

As can be seen in figure 11 this anisotropic shift $K_{ax}$ evidences a Curie–Weiss paramagnetism of the C$_{60}^{3-}$, which does not differ for the two isomers above 100K [20]. These data are compatible with a low spin $S$=1/2 effective moment as expected from figure 3 for a JT stabilized state. Furthermore the structural difference between the two isomers results in distinct magnetic ground states at $p$=1bar. The bipartite A15 structure displays a Néel order below $T_N$=47K [19,20,70], while the frustration of the fcc lattice results in a spin freezing only below 10K for the standard fcc- Cs$_3$C$_{60}$ [20,71].

Under an applied pressure, superconductivity is found to appear abruptly at different critical pressures $p_c$ for the two phases, with a simultaneous abrupt loss of magnetism, which definitely points toward a first order transition. The phase diagrams slightly differ for the two Cs$_3$C$_{60}$, but merge together [20,71] if plotted versus $V_{C60}$, the unit volume per C$_{60}$ ball (figure 12). There, the $T_c$ data can be scaled as well with that for the other known fcc-$A_3$C$_{60}$, and it has been evidenced that a similar maximum of $T_c$ versus $V_{C60}$ applies for the two structures [20,71,72].

Well above $p_c$, the SC state behaves as found for the non expanded fcc-phases, *e.g.* reductions of Knight shift and of $(T_1T)^{-1}$ are detected below $T_c$ for the two phases [20], so that overall the SC state does not seem to depend on the actual structure, except through the modification of the interball distance.

Above the magnetic and SC ground states, the x-ray and NMR spectra did not display any significant change with increasing pressure through the paramagnetic insulator to metal transition, for both isomers. This absence of structural modifications at $p_c$ establishes then that the evolution with $p$ only implies electronic degrees of freedom. Finally, $^{133}$Cs NMR shift and $T_1$ data also permitted to evidence the similar evolution of the electronic properties for the two isomers at the insulator metal transition [71].

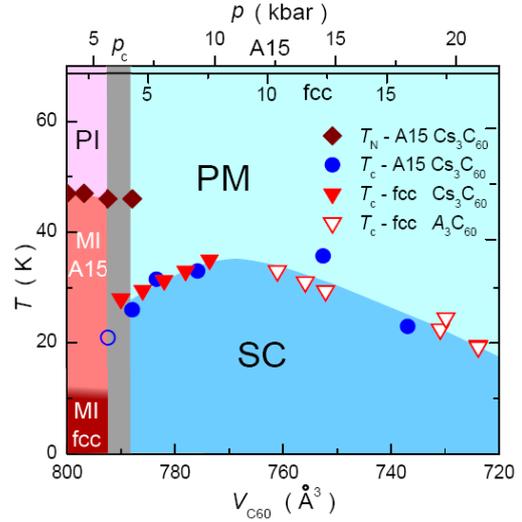

**Fig. 12.** Phase diagram representing $T_c$ versus the volume $V_{C60}$ per C$_{60}$ ball, and the Mott transition at $p_c$ (hatched bar) for the two isomer phases [71]. The corresponding pressure scales are shown on the upper scale. The transition to a magnetic insulating (MI) phase occurs only below 10 K in fcc-Cs$_3$C$_{60}$.

# 7 Discussion

We have recalled here a full set of experimental results which evidence the large incidence of Jahn-Teller effects on the physical properties of $A_n$C$_{60}$ compounds. Those help to establish a robust insulating state for evenly doped compounds, which can undergo in some cases a transition to a metallic state under pressure. We have evidenced that, in the $n$=3 cubic compounds the JT effects do help as well to maintain a metallic state, which remains however quite close to a Mott transition. The ground state of the metal is a singlet *s*-wave SC state in which pairing is driven by on ball phonons.

The expanded Cs$_3$C$_{60}$ compounds are unique inasmuch as they are insulating at $p$=1bar and can be driven through the Mott transition by an applied pressure. The phase diagram displays a SC dome, reminiscent of that found in many other correlated electron systems, with a $T_c$ maximum at a pressure somewhat above that of the first order Mott insulator state boundary.

In the SC state, the experiments established that the pure BCS equation or its Migdal-Eliashberg extension do not yield even a qualitative explanation of the deviations found near the Mott transition, where $T_c$ increases with pressure. There the normal state electronic properties deviate from Fermi liquid behaviour as spin fluctuations in the metallic state become prominent before the system switches into the Mott state [71]. Also, the orientational disorder of the C$_{60}$ in fcc-Cs$_3$C$_{60}$ does not appear detrimental to the *s*-wave SC, even near the Mott transition, as the $T_c$ dome does not differ from that of the A15 ordered isomer. This opposes the case of cuprates for which both normal state and *d*-wave SC are very sensitive to disorder [74-75].

On the theory side, serious efforts to include JT effects and correlations using DMFT have allowed to evidence that the Coulomb repulsion is at least not that detrimental to this on ball pairing [64]. It is important to recall here that ground state energy calculations done with DMFT have even allowed the anticipation [73] of the phase diagram with a SC dome near the Mott transition. This is a strong indication that detailed theoretical understanding of the properties of this series of materials is within reach. Further developments should then help to decide whether the large $T_c$ values in $A_3C_{60}$ compounds do result from a fundamental cooperation between correlations and electron-phonon interactions, and to determine above which pressure or bandwidth the BCS limit would be recovered.

Finally, we propose here that $Cs_3C_{60}$ is a rather good model system to study a 3D Mott transition. Indeed the latter is not controlled by doping, but is only driven by the inter-ball distance, that is the $t_{1u}$ bandwidth. In the (p,T) range probed [72] the qualitative behaviour detected appears quite similar to that expected for a single orbital Mott transition [76] although the $S=1/2$ ground sate has orbital degeneracy. Extensive measurements of other spectral and thermodynamic properties, possibly on larger $T$ ranges, are required to complete the experimental insight on this Mott transition.

# 8 Acknowledgments

The author acknowledges particularly V. Brouet who has been the major actor in the work on the non SC compounds and shows a continuing interest on the matter. I thank as well L. Forró and his group, with whom collaborations have been extending over many years. I should like also to acknowledge the expertise of P. Wzietek in the high pressure techniques which permitted us, with Y. Ihara, to develop the recent experiments on $Cs_3C_{60}$ compounds. Those were only made possible through the permanent exchanges with our collaborators M. Riccò, D. Pontiroli and M. Mazzani, who have been essential in the synthesis and x-ray characterization of the $Cs_3C_{60}$ compounds.

I also warmly thank E. Tosatti, M. Fabrizio and M Capone with whom we have had over the years many fruitful discussions about the theoretical aspects of the electronic structure of the fulleride compounds.